\documentclass[11pt,a4paper]{article}
\usepackage[utf8x]{inputenc}
\usepackage{jheppub}
\usepackage{youngtab}
\usepackage{hyperref}

\newdimen\tableauside\tableauside=1.0ex
\newdimen\tableaurule\tableaurule=0.4pt
\newdimen\tableaustep

\def\phantomhrule#1{\hbox{\vbox to0pt{\hrule height\tableaurule
width#1\vss}}}
\def\phantomvrule#1{\vbox{\hbox to0pt{\vrule width\tableaurule
height#1\hss}}}
\def\sqr{\vbox{%
 \phantomhrule\tableaustep
\hbox{\phantomvrule\tableaustep\kern\tableaustep\phantomvrule\tableaustep}%
 \hbox{\vbox{\phantomhrule\tableauside}\kern-\tableaurule}}}
\def\squares#1{\hbox{\count0=#1\noindent\loop\sqr
 \advance\count0 by-1 \ifnum\count0>0\repeat}}
\def\tableau#1{\vcenter{\offinterlineskip
 \tableaustep=\tableauside\advance\tableaustep by-\tableaurule
 \kern\normallineskip\hbox
   {\kern\normallineskip\vbox
     {\gettableau#1 0 }%
    \kern\normallineskip\kern\tableaurule}%
 \kern\normallineskip\kern\tableaurule}}
\def\gettableau#1 {\ifnum#1=0\let\next=\null\else
 \squares{#1}\let\next=\gettableau\fi\next}
\newcommand{\Yfund}{\tableau{1}}
\newcommand{\Ysymm}{\tableau{2}}
\newcommand{\Yasymm}{\tableau{1 1}}

\title{\center{Stringy Instantons in SU$(N) \;\mathcal N=2$ Non-Conformal Gauge Theories}}

\author{Hossein Ghorbani and}
\author{Daniele Musso}

\affiliation{Dipartimento di Fisica Teorica, Universita` di Torino and I.N.F.N., sezione di Torino, \\
Via P. Giuria 1, I-10125 Torino, Italy}

\emailAdd{ghorbani@to.infn.it}
\emailAdd{mussod@to.infn.it}

\abstract{In this paper we explicitly obtain the leading corrections to the SU$(N)\;\mathcal N=2$ prepotential 
due to stringy instantons both in flat space-time and in the presence of a non-trivial graviphoton background field.
We show that the stringy corrections to the prepotential are expressible in terms of the elementary symmetric polynomials. 
For $N>2$ the theory is not conformal; we discuss the introduction of an explicit dependence
on the string scale $\alpha'$ in the low-energy effective action through the stringy 
non-perturbative sector.
}


\begin{document}
\begin{flushright}
 DFTT/29/2011
 \end{flushright}

\maketitle 

\section{Introduction}

In the last fifteen years, the non-perturbative sector of supersymmetric gauge theories \cite{Dorey:2002ik,Bianchi:2007ft} has been
deeply investigated employing string theory techniques. 
The seminal papers \cite{Witten:1995im,Douglas:1995bn,Douglas:1996uz} firstly explored the connection between D-branes in string theory
with non-perturbative corrections in the corresponding low-energy effective field theories.
In fact, it has been shown that the ordinary field theory instanton calculus can be naturally and precisely rephrased 
in terms of D-brane models;
for instance, the ADHM construction yielding a parameterization of the instanton
moduli space can be read in terms of D-branes bound states (see for instance \cite{Billo:2002hm,Green:2000ke}). 

Several important results and explicit computations within the context of ordinary 
instanton calculus in gauge theory have been obtained using the string approach.
Let us just mention as an instance the detailed analysis and checks of Seiberg-Witten duality, and of string
dualities involving D-instantons \cite{Bruzzo:2002xf,Billo:2009di,Fucito:2009rs,Billo:2011uc}.
Moreover, in addition to the ordinary instanton calculus, 
the string theory framework opens the possibility of
investigating new features in non-perturbative physics.
This very interesting scenario consists in generalizing the ordinary instanton
models.
Indeed, it is possible to consider D-brane configurations which yield new kinds of non-perturbative corrections to the underlying effective 
gauge theories \cite{Billo':2010bd,Ghorbani:2010ks}.
Such generalized instantonic configurations are commonly referred to as \emph{exotic} or \emph{stringy} (for
a review see for instance \cite{Blumenhagen:2009qh}) since
their field theoretical interpretation is generally still an open question \cite{Billo:2009gc}.

The stringy instanton configurations present many peculiar and interesting features which are worth studying carefully.
Within the context of D-brane constructions, there is always a D-brane stack whose world-volume
contains the four-dimensional spacetime. 
We refer to these branes as the \emph{gauge} branes.
In our specific model, the world-volume of the gauge brane stack coincides with the
physical spacetime.
As opposed to the gauge branes, the \emph{instanton} branes are by definition point-like from 
the four-dimensional spacetime perspective\footnote{
In the D$3/$D$(-1)$ model under consideration the D$3$ branes represent tha gauge branes whereas
the D$(-1)$ branes are the instanton branes.
In other models, such as the D$7/$D$3$ systems, the D7 gauge branes
contain the $4$-dimensional spacetime and four extra dimensions which are wrapped in the internal space.}.

In a D-brane setup describing an ordinary instanton, the only difference between gauge and instanton branes
lies on the fact that while the gauge branes are extended along the $4$-dimensional spacetime directions, the instanton branes 
are here localized:
for ordinary configurations
the geometrical arrangement and transformation behavior in the internal space is identical between gauge 
and instanton branes.
The hallmark of the exotic character for an instanton configuration is instead 
represented by a different internal space geometry or symmetry properties 
between the gauge and the instanton branes.
In the D$7/$D$3$ systems leading to exotic instantons, the distinction is usually given by different wrappings of the gauge and 
the instanton branes in the internal space, namely a different geometrical arrangement. 
In the exotic configurations of our D$3/$D$(-1)$ model, instead, the internal geometry of the D$3$ and the D$(-1)$ branes is the same however
they behave differently under the orbifold action.
As we will see in Sec.\ref{dim_analysis}, the fact that instanton and gauge branes share the same arrangement 
in the internal space leads to an equal classical action for ordinary and exotic instantons.

In general, one of the main consequences of the different internal behavior between gauge and instanton branes is the absence of 
charged bosonic moduli associated with open string modes stretching between gauge and instanton branes.
For the ordinary instantons, these moduli are associated with the ``size'' of the instanton itself\footnote{Indeed,
in some cases it is possible to propose a field theoretical interpretation of the exotic configurations as the ``zero size'' limit of
ordinary instantons (see for instance \cite{Billo:2009gc})}.
In the exotic case under consideration, the charged bosonic moduli 
are actually projected out by the orbifold action (see \cite{Ghorbani:2010ks} for details).

One of the main features of exotic instantons consists in the 
explicit dependence of their effects on the string scale $\alpha'$.
This implies that we have a signal of stringy aspects within the field theoretical context through the
non-perturbative sector of the effective gauge theories.
Apart from the intrinsic theoretical interest, this offers a natural introduction of a new scale into the low-energy gauge 
theory, a feature particularly desirable in a phenomenological perspective.
Actually, stringy effects could lead to an accommodation of some naturalness questions posed by the parameters of phenomenological models
such as the see-saw parameters or the hierarchy of the Yukawa couplings involved in GUT models (see for example \cite{Blumenhagen:2009qh}
and references  therein).
The exotic instanton configurations are able to produce perturbatively prohibited effects like Majorana mass terms
and are possible ingredients of SUSY breaking models, particularly because the exotic configurations involve
the introduction of features like orientifold planes.

In this paper we generalize the results obtained for SU$(2)$ in \cite{Ghorbani:2010ks} to SU$(N)$ gauge theory.
In section \ref{setup} we provide a brief description of the D$3/$D$(-1)$ brane setup which we consider with special attention 
to the moduli content of the model. 
The D-brane arrangement and the orbifold/orientifold projections lead to a system whose low-energy regime
is described by an ${\cal N}=2$ supesymmetric gauge theory with matter transforming in the symmetric representation
of the gauge group.
As described in section \ref{dim_analysis}, the setup under consideration yields a conformal gauge theory when the number 
of ``colors'' is $N=2$ whereas conformality is lost for $N>2$. 
The explicit exotic contributions to the prepotential are computed in Section \ref{str_cor_pre} for the lowest values of
the instanton topological charge $k$. In Section \ref{flat} we study the limit of such corrections in the case of
flat background, i.e. vanishing VEV for the graviphoton field. 
We conclude in Section \ref{conclu} with final remarks and discuss the perspective for future investigation.
Eventually, the appendices \ref{appeA}, \ref{appeB}, \ref{appeC} contain usefuls formul\ae$\ $that have been used throughout the main text.

\section{Description of the setup}
\label{setup}

There are different models presenting exotic instanton configurations (see \cite{Blumenhagen:2009qh} and references therein).
In the present paper we use exactly the same setup and follow the same notation introduced in Ref. \cite{Ghorbani:2010ks};
there, it has been shown that exploiting Nekrasov's localization techniques \cite{Nekrasov:2002qd,Nekrasov:2003rj}
one is able to evaluate directly and explicitly the stringy corrections to the low-energy effective action  
for SU$(2)$ gauge theory.
The task we are presently committed to is the elaboration of the same model and its generalization to gauge groups
with rank higher than one, namely number of ``colors'' $N$ bigger that $2$.

\begin{figure}[h]
  \begin{center}
\vspace{10pt}
\hspace{-30pt}
\begin{picture}(0,0)%
\includegraphics{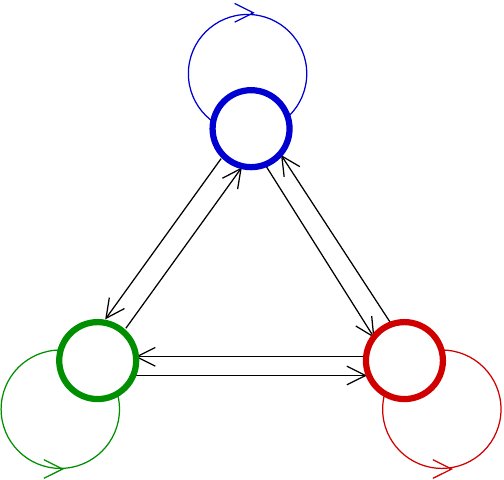}%
\end{picture}%
\setlength{\unitlength}{1450sp}%
\begingroup\makeatletter\ifx\SetFigFontNFSS\undefined%
\gdef\SetFigFont#1#2#3#4#5{%
  \reset@font\fontsize{#1}{#2pt}%
  \fontfamily{#3}\fontseries{#4}\fontshape{#5}%
  \selectfont}%
\fi\endgroup%
\begin{picture}(6030,5737)(541,-4460)
\put(1351,-2281){\makebox(0,0)[rb]{\smash{{\SetFigFont{12}{9.6}
 {\familydefault}{\mddefault}{\updefault}$\mathrm{U}(N_2)$}}}}
\put(2111,-2990){\makebox(0,0)[rb]{\smash{{\SetFigFont{10}{9.6}
 {\familydefault}{\mddefault}{\updefault}$N_2$}}}}
\put(5991,-171){\makebox(0,0)[rb]{\smash{{\SetFigFont{12}{9.6}
 {\familydefault}{\mddefault}{\updefault}$\mathrm{U}(N_1)$}}}}
\put(7491,-2281){\makebox(0,0)[rb]{\smash{{\SetFigFont{12}{9.6}
 {\familydefault}{\mddefault}{\updefault}$\mathrm{U}(N_3)$}}}}
\put(6111,-2990){\makebox(0,0)[rb]{\smash{{\SetFigFont{10}{9.6}
 {\familydefault}{\mddefault}{\updefault}$N_3$}}}}
\put(4111,-10){\makebox(0,0)[rb]{\smash{{\SetFigFont{10}{9.6}
 {\familydefault}{\mddefault}{\updefault}$N_1$}}}}
\end{picture}%
  \end{center}
\caption{The $\mathbb{C}^3/\mathbb{Z}_3$ orbifold model which corresponds to a configuration
of $N_1$, $N_2$ and $N_3$ fractional D3 branes before considering the orientifold projection. 
The arrows starting and ending on the
same node represent ${\cal N}=2$ vector multiplets in the adjoint representation of the $\mathrm{U}(N_i)$
groups. The arrows between different nodes represent bi-fundamental chiral multiplets
which pair up into ${\cal N}=2$ hypermultiplets.}
 \label{quiver}
 \end{figure}
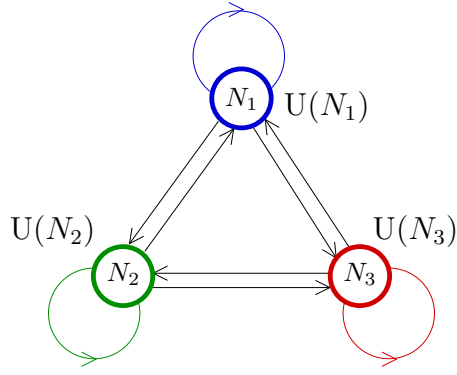

We focus the attention on a D$3/$D$(-1)$ model with $\mathbb{Z}_3$ orientifold 
background along the lines of Ref. \cite{Ghorbani:2010ks}. 
Let us here recall the main features of the employed brane model.
We consider a system of fractional D-branes, i.e. a set of branes placed at the singularity of the orbifold,
which carry different irreducible representations of the orbifold group as firstly discussed in \cite{Argurio:2007vqa}.
We choose the Chan-Paton factors of open strings that transform under the orbifold group 
according to the representations carried by the branes to which the string endpoints are attached.
The low-energy limit of the model is described by a gauge theory whose gauge group is encoded in the quiver diagram presented 
in Figure \ref{quiver}.

The extended directions of the D$3$ branes lie along the first four coordinates while the six internal directions
are parametrized with three complex variables $z^i$, $i=1,2,3$.
The orbifold acts on the space spanned by each complex variable $z^i$ with one of its irreducible representations 
according to:
\begin{equation}
 \label{gorb}
\begin{pmatrix}z^1 \cr  z^2 \cr z^3 \end{pmatrix}
~\to~
\begin{pmatrix}\xi\,z^1 \cr  \xi^{-1}\,z^2 \cr z^3 \end{pmatrix} \ ,
\end{equation}
\begin{table}[ht]
\begin{center}
\begin{tabular}{|c||c|c|c|}
\hline
\phantom{\vdots}
$(\Psi_0,\Psi_1)$&$\mathrm{SO}(k)$ &$\mathrm{SU}(N)$&$\mathrm{SU}(2)\times\mathrm{SU}(2)'$
\\
\hline\hline
$\phantom{\vdots}(a_\mu,M_\mu)$ & $\Ysymm$ &$\mathbf{1}$&$(\mathbf{2},\mathbf{2})$
\\
$\phantom{\vdots}(\lambda_c,D_c)$ & $\Yasymm$
 &$\mathbf{1}$&$(\mathbf{1},\mathbf{3})$
\\
$\phantom{\vdots}(\overline{\chi},\eta)$ &$\Yasymm$ & $\mathbf{1}$
 &$(\mathbf{1},\mathbf{1})$
\\
$\phantom{\Big|}(\mu,h)$ & $\Yfund$ &$\mathbf{\overline{N}}$&$(\mathbf{1},\mathbf{1})$
\\
$\phantom{\Big|}(\mu',h')$ & $\Yfund$ &$\mathbf{\overline{N}}$&$(\mathbf{1},\mathbf{1})$
\\
\hline
\end{tabular}
\end{center}
\caption{Moduli content of the stringy instanton configuration organized as BRST pairs and 
their transformation properties under the various symmetry groups.}
\label{tab:rep}
\end{table}
where $\xi=e^{\frac{2\text{i}\pi}{3}}$.
In addition, an O$3$ plane is added to the background so that its extended directions coincide with the extended directions of the D$3$ branes.
Before implementing the orientifold projection the theory has a hypermultiplet in the bi-fundamental representaion of the gauge groups associated to nodes 2 and 3 of the quiver diagram depicted 
in Fig. \ref{quiver}. 
It is not difficult to show that the orientifolding leads to an identification between these nodes and we end up with a hypermultiplet in the symmetric representation of the gauge group SU$(N)$ where 
$N\equiv N_1=N_2$. With this matter content the one-loop 
coefficient of the beta-function is $b_1=2-N$. Moreover, orientifolding reduces the instanton group from U$(k)$ to SO$(k)$.
Note that the representations of the orientifold acting on D-instantons and D$3$-branes are respectively symmetric and anti-symmetric
(see \cite{Ghorbani:2010ks}).
 
We define the following notation in order to account for the D-brane content:
$(N_1,N_2) \oplus (k_1,k_2)$ representing $N_i$ D$3$ and $k_i$ D$(-1)$ branes on the $i$-th node.
Notice that node $3$ does not appear because, as just stated, it is identified with node $2$.
We consider only configurations in which there are no D$3$-branes on node $1$, $N_1=0$; 
according to the results of \cite{Argurio:2007vqa,Ghorbani:2010ks}, a configuration presenting both D$3$ and D$(-1)$ branes on the node
$2$, $(0,N)\oplus(0,k)$, corresponds to an SU$(N)$ gauge theory with an ordinary instanton where $k$ represents the instanton number.
On the other hand, the configuration $(0,N)\oplus(k,0)$ is an SU$(N)$ gauge theory with an exotic $k$-instanton.
Note that the ordinary (exotic) character of the instantonic configurations is associated respectively with the
fact that the D$3$ and the D$(-1)$ branes are (are not) on the same node i.e. are (are not) associate to the same irreducible
representation of the orbifold group.
The details of the setup under considerations are already extensively illustrated in \cite{Ghorbani:2010ks},
here we only summarize the moduli spectrum in Table \ref{tab:rep}.

\section{Explicit dependence on the string scale and renormalization behavior of exotic effects}
\label{dim_analysis}

In the D-brane setups whose low-energy regime yields the ordinary $k$ instanton, the dimension of the integration measure
$d{\cal M}_k^{(\text{ord})}$ on the instanton moduli space 
is related to the one-loop coefficient $b_1$ of the $\beta$-function,
\begin{equation} \label{ord_dim}
 \left[d{\cal M}_k^{(\text{ord})}\right] = (\text{length})^{k b_1} \ .
\end{equation}
In order to have a dimensionless instanton action, we introduce a dimensionful prefactor to compensate
\eqref{ord_dim}.
This dimensionful coefficient in front of the ordinary $k$ instanton moduli integral is given by (see for instance \cite{Bianchi:2007ft}):
\begin{equation} \label{exp_ren_eq}
 \mu^{k b_1} e^{-\frac{8\pi^2}{g_{YM}^2}k} = \Lambda^{kb_1},
\end{equation}
where $\mu$ represents the high-energy renormalization mass scale and $\text{exp}(-8k\pi^2/g_{YM}^2)$ is the classical part of the instanton action.
In a stringy context, $\mu$ is naturally related to the string scale $\alpha'$,
\begin{equation} \label{ren_str_sc}
 \mu \sim \frac{1}{\sqrt{\alpha'}}.
\end{equation}
Note that the presence of $\mu$ introduces then an explicit dependence on the string scale.

Equation \eqref{exp_ren_eq} defines the dynamically generated infrared scale $\Lambda$
which in non-abelian gauge theory is indeed connected with non-perturbative effects.
Furthermore, it is crucial to observe that the dimensionful factor \eqref{exp_ren_eq}
associated with the ordinary instanton moduli integral can be expressed in terms of $\Lambda$ alone, 
that is to say, the dependence on $\alpha'$ disappears.
In other terms, the dependence on the high-energy renormalization scale is transmuted into
the low-energy scale $\Lambda$ which is completely interpretable in terms of the underlying effective field theory
description.

As already mentioned, stringy instantons present different features with respect to the ordinary cases. 
According to Ref. \cite{Ghorbani:2010ks}, the dimension of the moduli space in the $(0,N)\oplus(k,0)$ exotic
configuration is
\begin{equation}\label{exo_beta}
 \left[d{\cal M}_k^{(\text{ex})}\right] = (\text{length})^{k(2-N)} = (\text{length})^{-kb_1}.
\end{equation}
In Ref. \cite{Ghorbani:2010ks}
we specialized the analysis to the peculiar situation with SU$(2)$ gauge theory possessing 
the one-loop coefficient of the beta function $b_1$ equal to zero and being therefore conformal.
In the conformal case, even the exotic contributions lead to dimensionless non-perturbative corrections
which do not introduce explicitly the string scale $\alpha'$ in the low-energy effective theory.

Notice that in \eqref{exo_beta} the exponent has a different sign if compared with \eqref{ord_dim}. 
As a consequence, the dimensionful prefactor in front of the exotic moduli integral will be as follows:
\begin{equation} \label{exo_dim}
 \mu^{-kb_1} e^{-\frac{8\pi^2}{g_{YM}^2}k}.
\end{equation}
Two comments are in due here: The first observation is related to the classical part of the action which, in our model,
is the same for both ordinary and exotic configurations. The reason why this happens is described in detail in the end of this 
section.
The second point is that the dimensionful factor \eqref{exo_dim} in front of the exotic instanton moduli integral
cannot be expressed in terms of $\Lambda$ alone. 

Using the exponentiated renormalization group equation \eqref{exp_ren_eq} together with \eqref{ren_str_sc},
we have that the dimensionful exotic prefactor is proportional to
\begin{equation}
 (\alpha')^{kb_1} \Lambda^{kb_1}.
\end{equation}
As anticipated, the string scale is manifestly present in the exotic corrections to the prepotential
and couplings of the underlying low-energy gauge theory. 

We noted that the classical contribution of ordinary and exotic instantons within our model is the same.
This fact may seem unexpected but becomes natural as we observe that, in contrast to other possible brane models, in
our setup the gauge and the instanton branes have the same geometry in the internal space.
If this would not be the case, it is in general possible to have a different classical contribution for exotic and ordinary
configurations (see \cite{Blumenhagen:2009qh}).
Let us underline once more that in the model under consideration the feature marking the difference between ordinary
and the exotic instanton setups lies in the different transformation properties of the CP factors under the action of the orbifold.

\section{Stringy corrections to the prepotential}
\label{str_cor_pre}

\subsection{Preliminaries}
Nekrasov's localization technique\footnote{For details about Nekrasov's localization technique
and its applications see e.g. \cite{Nekrasov:2002qd,Nekrasov:2003rj}.} plays a crucial r\^ole in performing the explicit evaluation of
the partition function of the model under analysis.
The extended ${\cal N}=2$ supersymmetry of the system allows us to define a ``singlet'' linear combination of the supercharges
\begin{equation}
 Q' \equiv Q_{\dot{\alpha}\dot{\beta}} \epsilon^{\dot{\alpha}\dot{\beta}} \ ,
\end{equation}
representing a fermionic operator which can be shown to be nilpotent and interpretable as
a BRST operator.
Indeed, the action is $Q'$ exact i.e. it can be written as the $Q'$ variation of an appropriate fermionic expression $\Xi$ usually referred
to as the \emph{gauge fermion},
\begin{equation}
 S = Q'\, \Xi \ .
\end{equation}
The nilpotency of $Q'$ is consistent with the fact that the action $S$ is supersymmetric.
The BRST structure of the field content of the model allows us to rescale the fields and consider a particular limit known as
the \emph{localization limit} in which the action becomes quadratic in the fields.
The limit resembles formally a saddle-point approximation but it can be proven that, because of
the BRST structure of the model, it leads to exact results.

We are interested in computing the stringy non-perturbative partition functions corresponding to the lowest values of the instanton charge $k$. 
The total partition function is given by summing over all $k$-instanton 
partition functions as the following 
\begin{equation}\label{Zt}
 Z = \sum_{k=0}^\infty q^k Z_k \ ,
\end{equation}
where the dimensionful parameter $q$ is defined as
\begin{equation}\label{q}
 q \equiv \mu^{b_1} e^{2\pi \text{i} \tau}
\end{equation}
with $b_1$ the one-loop coefficient of the $\beta$-function. Notice that $b_1$ as seen in (\ref{exo_beta}) grows as the number of ``colors'' $N$ increases. Only for the case $N=2$ the coefficient $b_1$ 
vanishes and we deal with a conformal theory. In other cases, as discussed in Sec. \ref{dim_analysis} the partition function $Z_k$ has dimension $\mu^{-kb_1}$ which is compensated by the dimension 
of the prefactor $q^k$ yielding a dimensionless partition function $Z$.
The total prepotential is related to the partition function being proportional the logarithm of the function $Z$ (see e.g. \cite{Billo:2009di}),
\begin{equation}\label{Fnp}
 F^{(n.p.)}=\mathcal E \log Z
\end{equation}
where $\mathcal E$ as will be seen later, is the product of the eigenvalues of the graviphoton field.
Expanding the logarithm, the total prepotential is analogously obtained by adding up the contributions from all instanton charges, 

\begin{equation}\label{Fnp1}
 F^{(n.p.)}=\sum_{k=1}^\infty q^k F_k.
\end{equation}
The coefficients $F_k$'s are recursively obtained using equations (\ref{Zt}) and (\ref{Fnp}). The three leading coefficients which are of interest in this paper are given by
\begin{subequations}
 \begin{equation}\label{F1}
F_1=\mathcal E Z_1,
 \end{equation} 
\begin{equation}\label{F2}
F_2=\mathcal E Z_2-\frac{F_1^2}{2\mathcal E},\\   
 \end{equation} 
\begin{equation}\label{F3}
F_3=\mathcal E Z_3 - \frac{F_1 F_2}{\mathcal E} - \frac{F_1^3}{6\mathcal E^2}.  
 \end{equation}
\end{subequations}

Exploiting the localization techniques, the partition function $Z_k$ can be expressed as follows:
\begin{equation}
 Z_k=\mathcal N_k \int{\frac{d\chi}{2\pi \text{i}}\frac{\mathcal P(\chi) \mathcal R(\chi)}{\mathcal Q(\chi)}}\ .
\label{pf}
\end{equation}
The functions $\mathcal P(\chi)$, $\mathcal R(\chi)$ and $\mathcal Q(\chi)$ are given by
\begin{subequations}\label{prq}
\begin{align}
\mathcal P(\chi) \equiv \mathrm{Pf}_{\left(\tiny\Yvcentermath1{\yng(1,1)},\mathbf 1,\mathbf{(1,3)'}\right)}(Q'^2),\\
\mathcal R(\chi) \equiv \mathrm{det}_{\left(\tiny\yng(1),\mathbf{\bar N},\mathbf{(1,1)}\right)}(Q'^2),\\
\mathcal Q(\chi) \equiv \mathrm{det}^{1/2}_{\left(\tiny\Yvcentermath1\yng(2),\mathbf 1,\mathbf{(1,2)}\right)}(Q'^2).
\end{align}
\end{subequations}
Here we follow the same notation and conventions introduced in \cite{Ghorbani:2010ks} to which we refer for further details.
We simply recall that the subscripts in equations (\ref{prq}) denote the representations of the instanton group, 
gauge group and Lorentz group respectively. 
The action of the BRST charge squared $Q'^2$ on a generic modulus $\bullet$ is given by:
\begin{equation}
Q'^2 \bullet = T_{\text{SO}(k)}(\chi) \bullet -T_{\text{SU}(N)}(\phi) \bullet + T_{\text{SU}(2)\times \text{SU}(2)'}(\mathcal F)\ \bullet\ ,
\end{equation}
where the $T$'s are the infinitesimal transformations of the groups SO$(k)$, SU$(N)$ and SU$(2)\times$SU$(2)'$ in the appropriate representations. The matrices $\chi_{N\times N}$, $\phi_{k\times k}$
and $\mathcal F_{4\times 4}$ are respectively the gauge parameters of the groups above. Let us remind that $\phi$ is the scalar field belonging to the gauge sector 
(i.e. the open string states with both ends on the D3's), $\chi$ is the instanton D$(-1)$/$D(-1)$ modulus we have already introduced and $\mathcal F$ is the close string graviphoton field. 
The integral in (\ref{pf}) is clearly 
singular in correspondence to the zeros of the denominator $\mathcal Q(\chi)$. 
Moreover, the integrand becomes one in the limit $\chi \rightarrow \infty$;
this singularity is spurious and can be cured as described in \cite {Moore:1998et}. In the integral (\ref{pf}) the modulus $\chi$ has been assumed real.
By shifting the zeros of $\mathcal Q$ by a small positive value along the imaginary axis, say i$\epsilon$, the integration over $\chi$ 
can be treated as a contour integral in the complex upper half-plane. 
 This as well regularizes the divergence at infinity. 
Nevertheless, the direct evaluation of (\ref{pf})
is possible only by brute-force for just small instanton numbers and small gauge group ranks;  
we can however use all the symmetries of the theory in order to shorten the length of the calculations
and making it possible to extend the approach also to higher instanton numbers \cite{Ghorbani:2010ks} and higher gauge group ranks.

The partition function given in (\ref{pf}) is invariant under 
three symmetry groups: the instanton group SO$(k)$, the gauge group SU$(N)$ and the Lorentz group SU$(2)\times \text{SU}(2)'$. 
One can take advantage of these symmetries to express all the fields in the Cartan basis in which 
the integration becomes simpler. 
Changing the basis for the fields $\phi$ and $\mathcal F$ associated with the gauge group and the Lorentz group
respectively, causes no change in the integrand of (\ref{pf}), however expressing $\chi$ 
into the Cartan subalgebra is possible at a price of introducing a Vandermonde determinant, 
because the modulus $\chi$ appears in the integration measure;
\begin{equation}
\chi \rightarrow \overrightarrow\chi\cdot\overrightarrow H_{\text{SO}(k)}=\chi_i H^i_{\text{SO}(k)}
\end{equation}
$H^i_{\text{SO}(k)}$ being the Cartan generators of SO$(k)$ group and $i=1,...,\text{rank}\;\text{SO}(k)$.
The partition function then takes the following form
\begin{equation}
 Z_k=\mathcal N_k \int{\prod_i \frac{d\chi_i}{2\pi \text{i}} \Delta(\overrightarrow\chi) \frac{\mathcal P(\overrightarrow\chi) 
\mathcal R(\overrightarrow\chi)}{\mathcal Q(\overrightarrow\chi)}}\ ,
\label{pf1}
\end{equation}
where $\Delta(\chi)$ denotes the Vandermonde determinant. 
The functions in the integrand (\ref{pf1}) are related to a determinant or a Pfaffian of the operator $Q'^2$ 
in the appropriate representation to which the moduli belong. 
The values of such functions is given by the product of all 
eigenvalues of $Q'^2$ in the corresponding representations. 
Following \cite{Billo:2009di}, we exploit the weights of the representations to rewrite 
the functions appearing in the integrand \eqref{pf1}. 
The Vandermonde determinant is given by
\begin{equation}\label{D}
\Delta\left(\overrightarrow{\chi}\right)=\prod_{\overrightarrow{\rho}\in\, \text{adj}\neq0}\overrightarrow{\chi}.\overrightarrow{\rho},
\end{equation}
where $\overrightarrow{\rho}$ is the \emph{root vector} of SO$\left(k\right)$.
The VEV of the chiral superfield $\phi$, expressed in the Cartan basis
with $H_{\text{SU}\left(N\right)}^{u}$ the generators of the Cartan subalgebra of the SU$(N)$ group, is given by 
\begin{equation}
 \phi\rightarrow\overrightarrow{\phi} \cdot \overrightarrow{H}_{\text{SU}\left(N\right)}=\sum_{u}\phi_{u}H_{\text{SU}\left(N\right)}^{u}
\end{equation}
in which $u=1,...,\text{rank}\;\text{SU}(N)$. The function $\mathcal{R(\chi)}$ in the numerator of the integrand (\ref{pf1}) is expressed as

\begin{equation}\label{R}
 \mathcal{R}\left(\overrightarrow{\chi}\right)=\prod_{\overrightarrow{\pi}\in\boldsymbol{k}}\prod_{\overrightarrow{\gamma}\in N} 
\left(\overrightarrow{\chi} \cdot \overrightarrow{\pi}-\overrightarrow{\phi} \cdot \overrightarrow{\gamma}\right)
\end{equation}
where $\overrightarrow{\pi}$ and $\overrightarrow{\gamma}$ are the \emph{weight vectors}, respectively, in the vector and fundamental representations
of SO$\left(k\right)$ and SU$\left(N\right)$. Similarly,
\begin{equation}\label{P}
 \mathcal{P}\left(\overrightarrow{\chi}\right)=\prod_{\overrightarrow{\rho}\in\, \text{adj}}\prod_{\overrightarrow{\alpha}}^{\left(+\right)}
\left(\overrightarrow{\chi} \cdot \overrightarrow{\rho}-\overrightarrow{f} \cdot \overrightarrow{\alpha}\right)=
\prod_{\overrightarrow{\rho}\in\, \text{adj}}\prod_{a}\left(\overrightarrow{\chi} \cdot \overrightarrow{\rho}-f_{a}\right)\ .
\end{equation}
Here $\overrightarrow{\rho}$ is again the weight vector (root vector) in the adjoint representation of 
SO$\left(k\right)$ and $\overrightarrow{\alpha}$ as explained in \cite{Billo:2009di},
is the positive weight vector of the auxiliary SO$\left(3\right)$ group associated with the anti-selfdual/anti-chiral part of the
Lorentz group (i.e. the group rotating the $\lambda_{c}$'s). Therefore $a$ in \eqref{P} takes only the value $a=1$.
The graviphoton field which belongs to the Lorentz group SU$(2)\times$SU$(2)'$ 
along the Cartan directions takes the following form
\begin{equation}
\mathcal{F}=\left(\begin{array}{cc}
E_{1}\sigma_{2} & 0\\
0 & E_{2}\sigma_{2}\end{array}\right)\ .
\label{F}
\end{equation}
The function $Q(\overrightarrow\chi)$ in the denominator of the integrand (\ref{pf}) is expressed as
\begin{equation}\label{Q}
\mathcal{Q}\left(\overrightarrow{\chi}\right)=\prod_{\overrightarrow{\sigma}\in\, \text{sym}}\prod_{\overrightarrow{\beta}}^{\left(+\right)}
\left(\overrightarrow{\chi}.\overrightarrow{\sigma}-\overrightarrow{f}.\overrightarrow{\beta}\right)=\prod_{\overrightarrow{\sigma}\in
\, \text{sym}}
\prod_{A=1}^{2}\left(\overrightarrow{\chi}.\overrightarrow{\sigma}-E_{A}\right) 
\end{equation}
where $\overrightarrow{\sigma}$ denotes the weight vector in the symmetric representation of the
instanton group SO$(k)$ and $\overrightarrow{\beta}$ 
is the positive weight of the Lorentz group SU$(2)\times$SU$(2)'$ which can be read immediately from (\ref{F}).

\subsection{1-instanton contribution}
\label{k1}

In the $k = 1$ case, the instanton group is SO$(1)$ so the Cartan basis is $1$-dimensional and $\overrightarrow{\chi}$
has only one component $\chi$. 
Following equations \eqref{D}, \eqref{P}, \eqref{R}, \eqref{Q}, the Vandermonde determinant 
$\Delta(\chi)$ and the functions $\mathcal P(\chi)$, $\mathcal R(\chi)$, $\mathcal Q(\chi)$ are trivially given by 
\begin{eqnarray}
 &\Delta(\chi)&=1\ ,\\
 &\mathcal P(\chi)&=1\ ,\\
 &\mathcal R(\chi)&= \prod_{\overrightarrow{\gamma}\in N}\left(-\overrightarrow\Phi \cdot \overrightarrow\gamma\right)\ ,\\
 &\mathcal Q(\chi)&=E_1E_2=\mathcal E\ .
\end{eqnarray}
The partition function  \eqref{pf1} for $k=1$ then becomes
\begin{equation}
 Z_1=\frac{\mathcal N_1}{\mathcal E}\prod_{\overrightarrow{\gamma}\in N}\left(-\overrightarrow\Phi.\overrightarrow\gamma\right)\ ,
\end{equation}
where the $\overrightarrow\gamma$'s are the weights of SU$(N)$ in the fundamental representation. Using the explicit expression for the weights of the SU$(N)$ group, one obtains (see Appendix A)
\begin{equation}\label{detphi}
 \prod_{\overrightarrow{\gamma}\in{N}}\left(-\overrightarrow\Phi.\overrightarrow\gamma\right)=(-1)^N\det(\Phi)\ .
\end{equation}
The exotic 1-instanton partition function for SU$(N)$ gauge theory is therefore
\begin{equation}\label{Z_1}
 Z_1 = \frac{\mathcal N_1}{\mathcal E}(-1)^N \det(\Phi)\ ,
\end{equation}
where $\mathcal N_1$ is an overall normalization coefficient.

\subsection*{1-instanton prepotential \texorpdfstring{$F_1$}{}}

The exotic 1-instanton contribution to the prepotential is:
\begin{equation}
 F_1=\mathcal E Z_1\ ,
\end{equation}
therefore, using \eqref{Z_1}, we obtain
\begin{equation}\label{F_1}
 F_1 =\mathcal N_1 (−1)^N \det(\Phi)\ . 
\end{equation}

\subsection{2-instanton contribution}
\label{k2}

The $k=2$ partition function integral reads:
\begin{equation}
 Z_2 = (−1)^{N+1}\frac{\mathcal N_2 f}{\mathcal E}\int{\frac{d\chi}{2\pi \text{i}}} 
\frac{\prod_{i=1}^{N}\left[\chi^2-\left(\Phi \cdot \gamma_i\right)^2\right]}
{\prod_{A=1}^{2}(2\chi-E_A)(-2\chi-E_A)}
\end{equation}
Choosing the contour of integration to be in the upper half-plane, there are only two
positive poles, $\chi = \frac{E1}{2}$ and $\frac{E2}{2}$, enclosed in the integration path.
The result of the integration is simply given by the sum of the corresponding residues, namely
\begin{equation}\label{crude_res}
Z_{2}=\frac{\left(-1\right)^{N+1}\mathcal{N}_{2}}{4\mathcal E\left(E_{1}-E_{2}\right)}\left(\frac{\prod_{i=1}^{N}\left[\left(\frac{E_{1}}{2}\right)^{2}
-\left(\overrightarrow{\phi}\cdot\overrightarrow{\gamma_{i}}\right)^{2}\right]}{E_{1}}-\frac{\prod_{i=1}^{N}\left[\left(\frac{E_{2}}{2}\right)^{2}
-\left(\overrightarrow{\phi}\cdot\overrightarrow{\gamma_{i}}\right)^{2}\right]}{E_{2}}\right) \ ,
\end{equation}
where we have set $f = E_1 + E_2$.
However, this is not the desired form for the
result because in the limit of flat background, i.e. $E_1 , E_2 \rightarrow 0$, the prepotential is seemingly
divergent.
Moreover, we do not have yet a concrete answer in terms of the physical invariant
quantities, namely $\text{tr}\Phi^2$ and $\det\Phi$. 
The numerators in equation \eqref{crude_res} need therefore to be further elaborated. 
In fact, it turns out that the expansions of the polynomials in the numerators of
\eqref{crude_res} are just monomials where the coefficients are given in terms of the elementary symmetric polynomials (see appendix B).
Expanding equation \eqref{crude_res} we obtain
\[
Z_2=\frac{\mathcal{N}_{2}}{8\mathcal{E}\left(E_{1}-E_{2}\right)}\left\{ \sum_{j=0}^{N-2}\left(-1\right)^{N+j+1}e_{j}
\left(X_{1},...,X_{N}\right)\left[\left(\frac{E_{1}}{2}\right)^{2\left(N-j\right)-1}-\left(\frac{E_{2}}{2}\right)^{2\left(N-j\right)-1}\right]\right\}
\]
\begin{equation}\label{z_2}
+\frac{\mathcal{N}_{2}}{16\mathcal{E}}e_{N-1}\left(X_{1},...,X_{N}\right)
+\frac{\mathcal{N}_{2}}{4\mathcal{E}^{2}}e_{N}\left(X_{1},...,X_{N}\right)\ ,
\end{equation}
where the elementary symmetric polynomials $e_j(X_1 , ..., X_N)$ with 
$X_i=\left(\overrightarrow\Phi\cdot\overrightarrow\gamma_i\right)^2$ have been defined in Appendix B. 
The advantage of using elementary symmetric polynomials is that they can be translated into power sums 
which in turn are essentially polynomials in $\text{tr}\Phi^2$. 
From
(\ref{e_k}), (\ref{p_k}) and (\ref{tr^k}) one has
\begin{equation}
e_{j}\left(\left(\overrightarrow{\phi}\cdot\overrightarrow{\gamma_{1}}\right)^{2},...,\left(\overrightarrow{\phi}\cdot\overrightarrow{\gamma_{N}}\right)^{2}\right)=
\frac{1}{j!}\left|\begin{array}{ccccc}
\text{tr}\Phi^{2} & 1 & 0 & \cdots\\
\text{tr}\Phi^{4} & \text{tr}\Phi^{2} & 2 & 0 & \cdots\\
\vdots &  & \ddots & \ddots\\
\text{tr}\Phi^{2\left(j-1\right)} & \text{tr}\Phi^{2\left(j-2\right)} & \cdots & \text{tr}\Phi^{2} & j-1\\
\text{tr}\Phi^{2j} & \text{tr}\Phi^{2\left(j-1\right)} & \cdots & \text{tr}\Phi^{4} & \text{tr}\Phi^{2}
\end{array}\right|
\label{e_j}
\end{equation}
The expansion of the square brackets in equation (\ref{z_2}) making use of (\ref{x-y}) leads to
\[
\left(\frac{E_{1}}{2}\right)^{2\left(N-j\right)-1}-\left(\frac{E_{2}}{2}\right)^{2\left(N-j\right)-1}=
\left(\frac{1}{2}\right)^{2(N-j)}\left(E_{1}-E_{2}\right)\sum_{i=1}^{N-j-1}\text{tr}\mathcal{F}^{2\left(N-i-j\right)}\mathcal E^{i-1}
\]
\begin{equation}\label{expan}
+\left(\frac{1}{2}\right)^{2\left(N-j\right)-1}\left(E_{1}-E_{2}\right)\mathcal E^{N-j-1}
\end{equation}
where $\mathcal E = E_1 E_2$. From equation \eqref{F} the trace of the powers of the graviphoton field $\mathcal F$ reads
\begin{equation}\label{trF^2n}
\text{tr}(\mathcal{F}^{2n}) = 2\left(E_{1}^{2n}+E_{2}^{2n}\right)\ .
\end{equation}
Performing some algebraic passages exploiting (\ref{expan}) and (\ref{trF^2n}) one obtains the general 
partition function of the exotic 2-instanton in SU$(N)$ gauge theory in terms of the elementary symmetric polynomials $e_j$:
\begin{equation}
Z_{2}=\frac{\mathcal{N}_{2}}{4\mathcal{E}^{2}}e_N+\frac{\mathcal{N}_{2}}{16\mathcal{E}}e_{N-1}+
\sum_{j=0}^{N-2}\frac{\mathcal{N}_{2}\left(-1\right)^{N+j+1}}{2^{2\left(N-j\right)+3}}e_j
\left(\sum_{i=1}^{N-j-1}\text{tr}\mathcal{F}^{2\left(N-i-j\right)}\mathcal{E}^{i-2}+2\mathcal{E}^{N-j-2}\right)
\end{equation}
where $e_j$'s are given by (\ref{e_j}).
\subsection*{2-instanton prepotential $F_2$}

The $k = 2$ exotic contribution to the prepotential is given by
\begin{equation}\label{F_2}
 F_2=\mathcal E Z_2-\frac{F_1}{2 \mathcal E}\ ,
\end{equation}
where the $1$-instanton prepotential correction $F_1$ has been given in \eqref{F_1}.
Substituting $F_1$ and $Z_2$ in \eqref{F_2} we obtain
\[
F_{2}=\mathcal{N}_{2}\sum_{j=0}^{N-2}\frac{\left(-1\right)^{N+j+1}}{2^{2\left(N-j\right)+2}}e_{j}
\left(\sum_{i=1}^{N-j-1}\text{tr}\mathcal{F}^{2\left(N-i-j\right)}\mathcal{E}^{i-1}+2\mathcal{E}^{N-j-1}\right)
\]
\begin{equation}
+\frac{\mathcal{N}_{2}}{4\mathcal{E}}e_N-\frac{\mathcal{N}_{1}^{2}}{2\mathcal{E}}\det\Phi^2+\frac{\mathcal{N}_{2}}{16}e_{N-1}
\label{F_2_sum}
\end{equation}
It can be seen easily that none of the terms appearing in the summation \eqref{F_2_sum} is singular.
The only divergence arises in the second line and it is of order $\mathcal E^{−1}$ ,
\begin{equation}
 F_{2}=\frac{1}{\mathcal{E}}\left(\frac{\mathcal{N}_{2}}{4}-\frac{\mathcal{N}_{1}^{2}}{2}\right)\det\Phi^2+...
\end{equation}
where the dots indicate the non-divergent terms and we have used $e_N=\det\Phi^2$. 
So, in order to have no divergency, the free parameter $\mathcal N_2$ must 
satisfy the condition
\begin{equation}
 \mathcal N_2=2\,\mathcal N_1^2\ .
\end{equation}
Eventually, the prepotential for a $2$-instanton in the SU$(N)$ gauge theory is the following
\begin{equation}\label{gen_F_2}
F_{2}=\frac{\mathcal{N}_{1}^{2}}{8}e_{N-1}+\mathcal{N}_{1}^{2}\sum_{j=0}^{N-2}\frac{\left(-1\right)^{N+j+1}}{2^{2\left(N-j+1\right)}}e_{j}\left(
\sum_{i=1}^{N-j-1}\text{tr}\mathcal{F}^{2\left(N-i-j\right)}\mathcal{E}^{i-1}
+2\mathcal{E}^{N-j-1}\right)
\end{equation}
For the special case of SU$(2)$ gauge group, the exotic $k=2$ prepotential reads:
\begin{equation}
 F_{2}=-\frac{\mathcal{N}_{1}^{2}}{64}\left(
\text{tr}\mathcal{F}^{2}+2\mathcal{E}\right)
+\frac{\mathcal{N}_{1}^{2}}{8}e_{1}\ .
\end{equation}
The gauge field $\Phi$ in the Cartan basis of SU$(2)$ is given by the matrix
\begin{equation}
\Phi=\frac{1}{2}\left(\begin{array}{cc}
\phi & 0\\
0 & -\phi
\end{array}
\right),\ 
\label{phi}
\end{equation}
therefore $e_{1}=\text{tr}\Phi^2$ is equivalent to $e_1=-2\det\Phi$. 
The $k=2$ prepotential correction we read from the generic 
formula \eqref{gen_F_2} in the special case $N=2$ is therefore in complete agreement with the expression 
already obtained explicitly in \cite{Ghorbani:2010ks}.

\subsection{3-instanton contribution}
The evaluation of the exotic 3-instanton contribution to the prepotential
is still more involved than  the $1$ and $2$-instanton cases investigated in Secs. (\ref{k1}) and (\ref{k2}). 
Indeed in this case there are more poles in the path integral which make its evaluation lengthier. 
Some identities provided in Appendix C help to shorten the calculations.

The functions in the integrand of (\ref{pf}) can be given by (\ref{R}),(\ref{P}) and (\ref{Q}) for $k=3$.
The partition function then reads
\begin{equation}
Z_{3}=\left(-1\right)^{N}\frac{\mathcal{N}_{3}}{\mathcal E^{2}}\,\det\Phi\times I\ ,
\end{equation}
where 
\begin{equation}\label{I}
 I=\int\frac{d\chi}{2\pi \text{i}}\frac{f\chi^{2}\left(\chi-f\right)\left(\chi+f\right)\prod_{i=1}^{N}\left[\chi^{2}
-\left(\overrightarrow{\phi}\cdot\overrightarrow{\gamma_{i}}\right)^{2}\right]}{\prod_{A=1}^{2}\left(2\chi-E_{A}\right)
\left(-2\chi-E_{A}\right)\left(\chi-E_{A}\right)\left(-\chi-E_{A}\right)}
\end{equation}
Among the 8 poles in (\ref{I}) there are four positive poles, $\chi = E_1 , E_2 , E_1/2 , E_2/2$, that place 
inside the contour of integration encompassing the complex upper half-plane. 
The result of the integration is hence the sum of all the residues corresponding to the singularities inside the contour. 
We have
\begin{equation}
 I=I_{E_1}+I_{E_2}+I_{E_1/2}+I_{E_2/2}\ ,
\end{equation} 
where $I_{E_1},I_{E_2},I_{E_1/2},I_{E_2/2}$ are the residues at each pole given by:
\begin{equation}\label{IE1}
I_{E_1}=\frac{2E_{2}\prod_{i=1}^{N}\left[E_{1}^{2}-\left(\overrightarrow{\phi}\cdot\overrightarrow{\gamma_{i}}\right)^{2}\right]}{12E_{1}
\left(E_{2}-2E_{1}\right)\left(E_{1}-E_{2}\right)}
\end{equation}
\begin{equation}\label{IE2}
I_{E_{2}}=\frac{2E_{1}\prod_{i=1}^{N}\left[E_{2}^{2}-\left(\overrightarrow{\phi}\cdot\overrightarrow{\gamma_{i}}\right)^{2}\right]}{12E_{2}
\left(2E_{2}-E_{1}\right)\left(E_{1}-E_{2}\right)}
\end{equation}
\begin{equation}\label{IE3}
 I_{E_1/2}=\frac{\left(3E_{1}+2E_{2}\right)\prod_{i=1}^{N}\left[\left(\frac{E_{1}}{2}\right)^{2}
-\left(\overrightarrow{\phi}\cdot\overrightarrow{\gamma_{i}}\right)^{2}\right]}{12E_{1}\left(E_{1}-E_{2}\right)\left(E_{1}-2E_{2}\right)}
\end{equation}
\begin{equation}\label{IE4}
I_{E_{2}/2}=\frac{\left(3E_{2}+2E_{1}\right)\prod_{i=1}^{N}\left[\left(\frac{E_{2}}{2}\right)^{2}
-\left(\overrightarrow{\phi}\cdot\overrightarrow{\gamma_{i}}\right)^{2}\right]}{12E_{2}\left(E_{1}-E_{2}\right)\left(2E_{1}-E_{2}\right)}
\end{equation}
To obtain the residues above we have again set $f=E_1+E_2$. 
Taking advantage of the identity
\begin{equation}
\prod_{i=1}^{N}\left[\alpha^{2}-\left(\overrightarrow{\phi}.\overrightarrow{\gamma_{i}}\right)^{2}\right]=
\sum_{j=0}^{N}\left(-1\right)^{j}e_{j}\alpha^{2\left(N-j\right)}\ ,
\end{equation}
and performing some algebraic passages, one obtains the following expression for the integral \eqref{I}: 
\begin{equation}\label{I1}
I=\frac{\sum_{j=0}^{N}\left(-1\right)^{j}e_{j}\left\{ \left(E_{1}-2E_{2}\right)L_{N-j}\right.\left.+\left(E_{2}-2E_{1}\right)G_{N-j}\right\}}
{12E_{1}E_{2}\left(2E_{1}-E_{2}\right)\left(E_{1}-E_{2}\right)\left(E_{1}-2E_{2}\right)}\ ,
\end{equation}
where $L_k$ and $G_k$ are defined as
\begin{equation}\label{F_k}
L_k=2E_{2}^{2}E_{1}^{2k}-2E_{1}^{2}\frac{E_{2}^{2k}}{2^{2k}}-3E_{1}E_{2}\frac{E_{2}^{2k}}{2^{2k}}
\end{equation}
\begin{equation}\label{G_k}
G_k=3E_{1}E_{2}\frac{E_{1}^{2k}}{2^{2k}}+2E_{2}^{2}\frac{E_{1}^{2k}}{2^{2k}}-2E_{1}^{2}E_{2}^{2k} \ .
\end{equation}
In the presence of the graviphoton background field, the partition function $Z_k$ is expected to have the following form 
\begin{equation}\label{Z_k}
 Z_k=\mathcal E^{-k}f_k(\text{tr}\Phi^2,E_1,E_2)+\mathcal E^{-k+1}f_{k-1}(\text{tr}\Phi^2,E_1,E_2)+...+f_1(\text{tr}\Phi^2,E_1,E_2)\ ,
\end{equation}
where the $f_i$'s are regular polynomial functions of $\text{tr}\Phi^2$ and $E_1$, $E_2$. 
Again, it is clear that \eqref{I1} is not yet in the desired form. To go on further 
one should use the identities given in (\ref{FG0})-(\ref{FGN}). 
Employing these identities, the denominator of (\ref{I1}) cancels out an equal factor in the numerator.

After having expanded the numerator of (\ref{I1}) and having exploited (\ref{x-y}),
the partition function turns out to be:
\[
Z_3=\frac{\mathcal{N}_{3}}{12\mathcal{E}^{3}}\det\Phi e_{N}
+\frac{\mathcal{N}_{3}}{16\mathcal{E}^2}\det\Phi e_{N-1}
-\frac{\mathcal{N}_{3}}{192\mathcal{E}^2}\det\Phi e_{N-2}\left(\frac{3}{2}\text{tr}\mathcal{F}^{2}-5\mathcal{E}\right)
\]
\[
-\frac{\mathcal{N}_{3}}{\mathcal{E}^2}\det\Phi\sum_{j=0}^{N-3}\frac{\left(-1\right)^{N+j}e_{j}}{2^{2\left(N-j\right)+3}}\left(
\sum_{k=1}^{N-j-1}\text{tr}\mathcal{F}^{2\left(N-j-k\right)}\ensuremath{\mathcal{E}^{k-1}}+2\mathcal{E}^{N-j-1} \right)
\]
\[
+\frac{\mathcal{N}_{3}}{3\mathcal{E}^2}\det\Phi\sum_{j=0}^{N-3}\left(-1\right)^{N+j}e_{j}\sum_{i=1}^{N-j-2}\frac{1}{2^{i+2}}\left(
\sum_{k=1}^{N-j-i-1}\text{tr}\mathcal{F}^{2\left(N-j-i-k\right)}\ensuremath{\mathcal{E}^{k+i-1}}+2\mathcal{E}^{N-j-1}\right)
\]
\[
-\frac{\mathcal{N}_{3}}{3\mathcal{E}^2}\det\Phi\sum_{j=0}^{N-3}\left(-1\right)^{N+j}e_{j}\sum_{i=N-j+1}^{2\left(N-j-1\right)}\frac{1}{2^{i+2}}\left(
\sum_{k=1}^{i-N+j}\text{tr}\mathcal{F}^{2\left(i-N+j-k+1\right)}\ensuremath{\mathcal{E}^{2\left(N-j-1\right)+k-i}}+2\mathcal{E}^{N-j-1}\right)
\]
\begin{equation}\label{Z_3}
+\frac{\mathcal{N}_{3}}{3\mathcal{E}^2}\det\Phi\sum_{j=0}^{N-3}\left(-1\right)^{N+j}e_{j}\frac{1}{2^{N-j+1}}\mathcal{E}^{N-j-1}
\end{equation}
where the elementary symmetric polynomials $e_j$ have been defined in (\ref{e_j}) and $\text{tr}\mathcal F^{2n}$ is given in \eqref{F}.

\subsection*{3-instanton prepotential $F_3$}
The prepotential for the $k=3$ exotic instanton is given by
\begin{equation}\label{F_3}
F_{3}=\mathcal{E}Z_{3}-\frac{F_{2}F_{1}}{\mathcal{E}}-\frac{F_{1}^{3}}{6\mathcal{E}^{2}}\ .
\end{equation}
Plugging the partition function \eqref{Z_3}, the $2$ and $1$-instanton prepotentials \eqref{F_2} and \eqref{F_1}
into \eqref{F_3} we obtain 
\[
F_{3}=\left(\frac{\mathcal{N}_{3}}{12\mathcal{E}^{2}}-\frac{\mathcal{N}_{1}^{3}}{6\mathcal{E}^{2}}\right)\det\Phi^{3}
\]
\[
+\left(\frac{\mathcal{N}_{3}}{16\mathcal{E}}-\frac{\mathcal{N}_{1}^{3}}{8\mathcal{E}}\right)\det\Phi e_{N-1}
\]
\[
-\left(\frac{\mathcal{N}_{3}}{192\mathcal{E}}-\frac{\mathcal{N}_{1}^{3}}{64\mathcal{E}}\right)\det\Phi e_{N-2}\text{tr}\mathcal{F}^{2}
\]
\[
+\left(\frac{\mathcal{N}_{3}}{2\mathcal{E}}-\frac{\mathcal{N}_{1}^{3}}{\mathcal{E}}\right)\det\Phi\sum_{j=0}^{N-3}
\frac{\left(-1\right)^{N+j+1}}{2^{2\left(N-j\right)+1}}e_{j}\mathcal{E}^{N-j-1}
\]
\begin{equation}
+\left(\frac{\mathcal{N}_{3}}{2\mathcal{E}}-\frac{\mathcal{N}_{1}^{3}}{\mathcal{E}}\right)\det\Phi\sum_{j=0}^{N-3}
\frac{\left(-1\right)^{N+j+1}}{2^{2\left(N-j+1\right)}}e_{j}\sum_{i=1}^{N-j-1}\text{tr}\mathcal{F}^{2\left(N-i-j\right)}\mathcal{E}^{i-1}+...
\end{equation}
where the dots indicate the non-singular terms. 
We see that one single condition on the only free parameter $\mathcal N_3$ can entirely remove all the singularities; 
assuming that
\begin{equation}
  \mathcal{N}_{3}=2\mathcal{N}_{1}^{3}
\end{equation}
the 3-instanton prepotential in the SU$(N)$ gauge theory under consideration is
\[
F_{3}=\frac{\mathcal{N}_{1}^{3}}{12}\det\Phi\, e_{N-2}
\]
\[
+\frac{\mathcal{N}_{1}^{3}}{3}\det\Phi\,\sum_{j=0}^{N-3}\left(-1\right)^{N+j}e_{j}\mathcal{E}^{N-j-2}\left(1+\frac{1}{2^{2(N-j-1)}}-\frac{1}{2^{N-j-2}}\right)
\]
\[
+\frac{\mathcal{N}_{1}^{3}}{3}\det\Phi\,\sum_{j=0}^{N-3}\left(-1\right)^{N+j}e_{j}\sum_{i=1}^{N-j-2}\frac{1}{2^{i+1}}\mathcal{E}^{i-1}\sum_{k=1}^{N-j-i-1}
\text{tr}\mathcal{F}^{2\left(N-j-i-k\right)}\mathcal{E}^{k-1}
\]
\begin{equation}\label{prep3}
-\frac{\mathcal{N}_{1}^{3}}{3}\det\Phi\,\sum_{j=0}^{N-3}\left(-1\right)^{N+j}e_{j}\sum_{i=N-j+1}^{2\left(N-j-1\right)}\frac{1}{2^{i+1}}\mathcal{E}^{2\left(N-j\right)-i-2}
\sum_{k=1}^{i-N+j}\text{tr}\mathcal{F}^{2\left(i-N+j-k+1\right)}\mathcal{E}^{k-1}.
\end{equation}
As already done for lower instanton numbers, we can check the $k=3$ general formula against the special $N=2$ case explicitly
obtained in \cite{Ghorbani:2010ks}. 
The $F_3$ prepotential for the SU$(2)$ gauge theory is given by only the first line of \eqref{prep3}
because the second and the third lines are summations beginning from $N=3$, so   
\begin{equation}
 F_{3}=\frac{\mathcal{N}_{1}^{3}}{12}\det\Phi\ ,
\end{equation}
where we have used $e_0=1$. This coincides exactly with the result for $F_3$ provided in \cite{Ghorbani:2010ks}.

\section{Prepotential in flat background}
\label{flat}

The RR graviphoton background field we considered 
was a necessary tool in order to calculate the prepotential using Nekrasov's approach. 
It indeed regularizes the divergence due to the infinity coming from the integration over 
the moduli describing the position of the instanton \cite{Billo:2009di}.
Applying the localization method, one can retain just the quadratic terms in the action being
therefore able to perform explicitly the integration with no difficulty. 
In addition, after having obtained
the instanton contributions to the prepotential in the presence of the graviphoton background field one may be interested 
in the same contributions but in a flat background. To this end it is enough to let 
\begin{equation}\label{fb}
E_1, E_2 \rightarrow 0 \hspace{.5cm} \text{or equivalently} \hspace{.5cm} \mathcal F \rightarrow 0. 
\end{equation}
In this limit the $1$-instanton contribution remains unchanged
because it is actually independent of the background and the rank of the gauge group up to an overall sign. 
So in the flat background case we still have:
\begin{equation}
 F_1=(-1)^N\mathcal N_1 \det\Phi\ .
\end{equation}

For the $2$-instanton case the only non-vanishing term in (\ref{gen_F_2}) in the limit (\ref{fb}) turns out to be for $j = N − 1$ 
and $i = 1$. Therefore
\begin{equation}
F_{2}=\frac{\mathcal{N}_{1}^{2}}{8}e_{N-1}.
\end{equation}
Similarly the $3$-instanton prepotential in flat background results in
\begin{equation}
F_{3}=\frac{\mathcal{N}_{1}^{3}}{12}\det\Phi\, e_{N-2},
\end{equation}
where $e_j$ is given by (\ref{e_j}) and $\det\Phi$ has been introduced in terms of the weight vectors in Appendix \ref{appeA} .

\section{Final Remarks and Conclusion}
\label{conclu}

In the present paper we have found explicitly the non-perturbative stringy corrections 
to the $\mathcal N=2$ prepotential in SU$(N)$ gauge theory for generic $N$, thus
generalizing the results already obtained in \cite{Ghorbani:2010ks} for the
special case of SU$(2)$ gauge group.
The setup of the model is the same as the one described 
in \cite{Ghorbani:2010ks}, i.e. we take some set of D$3$-branes along with D-instantons (D$(-1)$-branes) in the 
orbifold $\mathbb C^2/\mathbb Z_3$ and O$3$-plane backgrounds. The stringy effects arise when D-instantons 
and D$3$-branes transform under different representations of the orbifold group. In other words, we deal with a 
triangular quiver diagram with the $(k,0)\oplus(0,N)$ arrangement of branes
appropriately chosen in order to obtain the desired low-energy effective theory.
So the exotic character of the instantons stems from the fact that the D$3$'s and the D$(-1)$'s
occupy different nodes of the quiver diagram.

For $N>2$ the one-loop coefficient $b_1$ of the $\beta$-function, takes non-zero values and
the corresponding gauge theory is therefore not conformal.
It was argued in Sec. \ref{dim_analysis} that, in the model considered in this paper, the exotic instanton moduli integral has dimension (length)$^{k(2-N)}$.
The partition function $Z_k$ of the $k$-instanton becomes therefore dimensionful. However, the total partition function $Z$ which is the sum over all instanton 
contributions as seen in (\ref{Zt}), remains dimensionless because of the introduction of a dimensionful factor in $q$ defined in (\ref{q}).

The prefactor $q^k$'s in (\ref{Fnp1}) compensate appropriately also the dimension of the $k$-instanton prepotential leading to the total prepotential with dimension (length)$^2$. It is worth 
noting that only for the special case $N=2$ all $k$-instanton prepotentials have the same dimension therefore they may be summed up to a logarithmic closed form as proposed 
in \cite{Ghorbani:2010ks}. For cases $N>2$ this does not happen 
anymore because all prepotentials associated to different instanton charges in the series (\ref{Fnp}) have different dimensions, hence they are no more additive.

One can check explicitly that the dimensionful quantities in (\ref{F_1}), (\ref{gen_F_2}) and (\ref{prep3}) compensate 
the dimension of the prefactor
yielding the correct dimension (length)$^2$ for the prepotential.
Let us elaborate more this point by looking for instance at the $k=2$ result (\ref{gen_F_2}). 
The dimensionful quantities appearing in $F_2$ are 
the elementary symmetric polynomials $e_j$'s, the trace of the graviphoton field tr$\mathcal F$, and $\mathcal E$. 
Noting that $E_1$ and $E_2$ have the dimension of length, the $\mathcal E$ and tr$\mathcal F$ both 
have dimension (length)$^{2}$, furthermore from (\ref{e_j}) can be seen that $e_j$ has dimension (length)$^{2j}$.
Considering all dimensionful terms in (\ref{gen_F_2}), it turns out that the dimension of $F_2$ equals (length)$^{2(N-1)}$ 
which combining with the dimension of the prefactor (mass)$^{2(N-2)}$ leads to the expected dimension (length)$^2$. The case $k=3$ can be checked similarly.

In order to compute the non-perturbative corrections due to the stringy instanton charges up to $k=3$,
we take advantage of the properties of the elementary symmetric polynomials and their relations 
with the power sums.
The general results that we find are 
perfectly in agreement with the special case $N=2$ provided in \cite{Ghorbani:2010ks}. 
In the limit of flat background (vanishing graviphoton background), 
we have shown that the results are simplified significantly 
because they are expressible using only the elementary symmetric polynomials.
Dimensional analysis of the measure of the moduli integral shows that for the
brane arrangement in the quiver diagram considered in this paper, the conformality of the gauge theory 
occures only if the gauge group is taken to be SU$(2)$.  
For higher gauge group ranks a dimensionful prefactor in front of the moduli integral 
is required in order to compensate the dimension of the moduli measure. 
The stringy character of the corrections is manifest in the prefactor
since it depends explicitly on the string scale $\alpha'$.
It may be interesting to investigate models in which we can look for stringy instantons in conformal theories for arbitrary gauge group ranks. 
The stringy corrections in the prepotential for such models would be independent of the string scale. One therefore may expect to have once again a logarithmic behavior for the total 
prepotential analogous to the logarithmic function proposed in \cite{Ghorbani:2010ks}.

\begin{acknowledgments}
 We would like to thank A. Lerda for his essential guidance and precious advice in (but not only) the preparation of this paper.
We would like to thank also Marco Bill\'o, Igor Pesando, Marialuisa Frau for several fruitful discussions. H.G. is also grateful for the hospitality
of the institute of theoretical physics in Vienna (Austria) university of technology where the last stages of this work were done. 
\end{acknowledgments}

\appendix
\section{Weights of SU\texorpdfstring{$(N)$}{} Fundamental Representation}
\label{appeA}

The rank of the group SU$(N)$ i.e. the dimension of the maximally commuting subalgebra
of $\mathfrak{su}(N)$ is $N − 1$. This means among all $N^2 − 1$ traceless hermitian generators, $N − 1$
generators (constructing Cartan subalgebra) are diagonalized. We denote the $N − 1$ Cartan
generators of SU$(N)$ group by $H_m $ where $m = 1, ..., N − 1$. These are $N \times N$ hermitian
and traceless diagonal real matrices that can be chosen to be %\cite{}.
\begin{equation}
\left[H_{m}\right]_{ij}=\frac{1}{\sqrt{2m\left(m+1\right)}}\left(\sum_{k=1}^{m}\delta_{ik}\delta_{jk}-m\,\delta_{i,m+1}\delta_{j,m+1}\right)
\end{equation}
A state in the fundamental representation of SU$(N)$ group is $N$ dimensional. There are
then equal number of $(N − 1)$-dimensional weight vectors defined by
\begin{equation}
\left[\gamma_{i}\right]_{m}=\left[H_{m}\right]_{ii}=\frac{1}{\sqrt{2m\left(m+1\right)}}\left(\sum_{k=1}^{m}\delta_{ik}-m\delta_{i,m+1}\right)
\end{equation}
where $i$ runs from $1$ to $N $ and denotes the $i$th weight. The index $m$ denotes the $m$th entry
of the $(N − 1)$-dimensional vector. The VEV of the super chiral field $\Phi$ as mentioned in
the text can be brought into the Cartan basis. It is therefore the linear combination of the
Cartan generators:
\begin{equation}
 \Phi=\sum_{m} \phi_m H_m
\end{equation}
now
\begin{equation}
 \det\Phi=\det{\left(\sum_{m} \phi_m H_m\right)}=\prod_i \sum_m \phi_m \left[H_m\right]_{ii}=\prod_i \sum_m \phi_m \left[\gamma_i\right]_{m}
=\prod_i \overrightarrow\Phi.\overrightarrow\gamma_i
\end{equation}
this proves (\ref{detphi}). We also have
\begin{equation}
 \text{tr}\Phi=\text{tr}{\left(\sum_{m} \phi_m H_m\right)}=\sum_i \left(\sum_m \phi_m \left[H_m\right]_{ii}\right)=\sum_i \left(\sum_m \phi_m \left[\gamma_i\right]_{m}\right)
=\sum_i \overrightarrow\Phi.\overrightarrow\gamma_i
\end{equation}
which implies
\begin{equation}\label{tr^k}
 \text{tr}\Phi^k=\sum_i \left(\overrightarrow\Phi.\overrightarrow\gamma_i\right)^k
\end{equation}

\section{Elementary Symmetric Polynomials and Power Sums}
\label{appeB}

The numerator in equations (\ref{crude_res}) and (\ref{IE1})-(\ref{IE4}) can be expanded into a monic polynomial; in general we
have
\begin{equation}
\prod_{j=1}^{N}\left(\lambda-X_{j}\right)=\lambda^{N}-e_{1}\left(X_{1},...,X_{N}\right)\lambda^{N-1}+e_{2}\left(X_{1},...,X_{N}\right)\lambda^{N-2}-...+\left(-1\right)^{N}e_{N}\left(X_{1},...,X_{N}\right)
\end{equation}
where $e_k (X_1 , ..., X_N )$ is called Elementary Symmetric Polynomial of order $k$ and is defined by
\begin{equation}
e_{k}\left(X_{1},...,X_{N}\right)=\sum_{1\leq j_{1}\leq j_{2}\leq...\leq j_{N}\leq N}X_{j_{1}}X_{j_{2}}...X_{j_{N}}
\end{equation}
which means
\begin{equation}
e_{0}\left(X_{1},...,X_{N}\right)=1
\end{equation}
\begin{equation}
e_{1}\left(X_{1},...,X_{N}\right)=\sum_{1\leq j\leq N}X_{j}
\end{equation}
\begin{equation}
 e_{2}\left(X_{1},...,X_{N}\right)=\sum_{1\leq j_{1}\leq j_{2}\leq N}X_{j_{1}}X_{j_{2}}
\end{equation}
\[
 \vdots
\]
\begin{equation}
e_{N}\left(X_{1},...,X_{N}\right)=X_1...X_N.
\end{equation}
When $X_1 , .., X_N$ are entries of a diagonalized matrix $X$ (which is so in this paper), then
\begin{equation}
e_{N}\left(X_{1},...,X_{N}\right)=\det X.
\end{equation}
For example for $N = 3$
\begin{equation}
e_{0}\left(X_{1},X_{2},X_{3}\right)=1
\end{equation}
\begin{equation}
e_{1}\left(X_{1},X_{2},X_{3}\right)=X_{1}+X_{2}+X_{3}
\end{equation}
\begin{equation}
e_{2}\left(X_{1},X_{2},X_{3}\right)=X_{1}X_{2}+X_{2}X_{3}+X_{2}X_{3}
\end{equation}
\begin{equation}
 e_{3}\left(X_{1},...,X_{3}\right)=X_1 X_2 X_3
\end{equation}
On the other hand the Power Sum of order $k$, $p_k (X_1 , ..., X_N )$ is defined by
\begin{equation}\label{p_k}
p_{k}\left(X_{1},...,X_{N}\right)=\sum_{i=1}^{N}X_{i}^{k}=X_{1}^{k}+...+X_{N}^{k}
\end{equation}
The relation between the Power Sum and the Elementary Symmetric Polynomials is given
by %\cite{}
\begin{equation}\label{e_k}
e_{k}=\frac{1}{k!}\left|\begin{array}{ccccc}
p_{1} & 1 & 0 & \cdots\\
p_{2} & p_{1} & 2 & 0 & \cdots\\
\vdots &  & \ddots & \ddots\\
p_{k-1} & p_{k-2} & \cdots & p_{1} & k-1\\
p_{k} & p_{k-1} & \cdots & p_{2} & p_{1}
\end{array}\right|
\end{equation}

\section{Useful Identities}
\label{appeC}

The functions $L_k$ and $G_k$ 
\begin{equation}
L_k=2E_{2}^{2}E_{1}^{2k}-2E_{1}^{2}\frac{E_{2}^{2k}}{2^{2k}}-3E_{1}E_{2}\frac{E_{2}^{2k}}{2^{2k}}
\end{equation}
\begin{equation}
 G_k=3E_{1}E_{2}\frac{E_{1}^{2k}}{2^{2k}}+2E_{2}^{2}\frac{E_{1}^{2k}}{2^{2k}}-2E_{1}^{2}E_{2}^{2k}
\end{equation}
introduced in (\ref{F_k}) and (\ref{G_k}) satisfy the following identities:
\begin{equation}\label{FG0}
\left(E_{1}-2E_{2}\right)L_0+\left(E_{2}-2E_{1}\right)G_0=\left(E_{1}-2E_{2}\right)\left(E_{1}-E_{2}\right)\left(2E_{1}-E_{2}\right)
\end{equation}
\begin{equation}\label{FG1}
\left(E_{1}-2E_{2}\right)L_1+\left(E_{2}-2E_{1}\right)G_1=-\frac{3}{4}E_{1}E_{2}\left(E_{1}-2E_{2}\right)\left(E_{1}-E_{2}\right)\left(2E_{1}-E_{2}\right)
\end{equation}
\[
\left(E_{1}-2E_{2}\right)L_2+\left(E_{2}-2E_{1}\right)G_2=
\]
\begin{equation}\label{FG2}
-\frac{1}{16}E_{1}E_{2}\left(E_{1}-2E_{2}\right)\left(E_{1}-E_{2}\right)\left(2E_{1}-E_{2}\right)\left[3\left(E_{1}^{2}+E_{2}^{2}\right)-5E_{1}E_{2}\right]
\end{equation}
and for $N-j\geq3$ we have
\[
\left(E_{1}-2E_{2}\right)L_{N-j}+\left(E_{2}-2E_{1}\right)G_{N-j}=
\]
\[
\frac{3}{2^{2\left(N-j\right)}}E_{1}E_{2}\left(2E_{1}-E_{2}\right)\left(2E_{2}-E_{1}\right)\left[E_{1}^{2\left(N-j\right)-1}-E_{2}^{2\left(N-j\right)-1}\right]
\]
\[
+\sum_{i=1}^{N-j-2}\frac{1}{2^{i-1}}\left(E_{1}E_{2}\right)^{i+1}\left(2E_{1}-E_{2}\right)\left(E_{1}-2E_{2}\right)\left[E_{1}^{2\left(N-j-i\right)-1}-E_{2}^{2\left(N-j-i\right)-1}\right]
\]
\[
-\sum_{i=N-j+1}^{2\left(N-j-1\right)}\frac{1}{2^{i-1}}\left(E_{1}E_{2}\right)^{2\left(N-j\right)-i}\left(2E_{1}-E_{2}\right)\left(E_{1}-2E_{2}\right)\left[E_{1}^{2\left(i-N+j\right)+1}-E_{2}^{2\left(i-N+j\right)+1}\right]
\]
\begin{equation}\label{FGN}
+\frac{1}{2^{N-j-1}}\left(E_{1}E_{2}\right)^{\left(N-j\right)}\left(2E_{1}-E_{2}\right)\left(E_{1}-2E_{2}\right)\left(E_{1}-E_{2}\right)
\end{equation}\\
We have also made use repeatedly of the following identity
\begin{equation}\label{x-y}
x^{2n-1}-y^{2n-1}=\left(x-y\right)\sum_{i=1}^{n-1}\left(x^{2\left(n-i\right)}+y^{2\left(n-i\right)}\right)
\ensuremath{\left(xy\right)^{i-1}+\left(x-y\right)\left(xy\right)^{n-1}}
\end{equation}

\bibliographystyle{kp}
\bibliography{gen}
\end{document}